\documentstyle[12pt]{article}
\begin{document}
\newcommand{\eqnref}[1]{Eq.~(\ref{#1})}
\begin{center}
{\bf Quantum Field Theory of the Laser Acceleration}\\[0.2cm]
Miroslav Pardy\\[0.2cm]
{Department of Theoretical Physics and Astrophysics, \\
Masaryk University, Faculty of Science, Kotl\'a\v rsk\'a 2, 611 37 Brno,\\
Czech Republic} \\
{e-mail:pamir@physics.muni.cz}
\end{center}

\vskip 2cm

\begin{abstract}
After the historical background concerning the pressure of
light, we derive the quantum field theory  force of
the laser radiation acting on electron. Numerically, we
determine the velocity of an electron accelerated by laser beam,
after acceleration time $\Delta t = 1 {\rm s}$.
\end{abstract}

\vskip 10mm

\baselineskip 15 pt

\section{Why laser acceleration ?}

The problem of acceleration of charged particles by the laser field is,
at present time, one of the most prestigeous problem in the accelerator
physics. It is supposed that, in the future, the laser accelerator will
play the same role in particle physics as the linear or circle accelerators
working in today particle laboratories.

The acceleration effectiveness of the linear or circle accelerators is
limitied not only by geometrical size of them
but also by the energy loss of accelerated particles
which is caused by bremsstrahlung during the
acceleration. The amount of radiation, following from the Larmor formula,
emitted by accelerated
charged particle is given generally as follows (Maier, 1991):

\begin{equation}
P = \frac {2}{3}\frac {r_{0}m}{c}(\gamma^{6}a_{||}^{2} + \gamma^{4}
a_{\perp}^{2}); \quad \gamma = \frac {1}{\sqrt{1-\beta^{2}}};\quad
\beta = v/c
\label{1}
\end{equation}
where $v$ is the velocity of a particle, $c$ is the velocity of light
in vacuum, $a_{||}$ is parallel acceleration of a particle and
$a_{\perp}$ is the perpendicular acceleration of a particle in
the accelerator, $m$ is the rest mass of an electron. The quantity

\begin{equation}
r_{0} = \frac {1}{4\pi \varepsilon_{0}}\frac {e^{2}}{m c^{2}}
\label{2}
\end{equation}
is the electron classical radius in SI units.

In terms of momenta

\begin{equation}
\dot p_{||} = \frac {\dot E}{\beta c}; \quad  \dot p_{\perp}  = m
\gamma \dot v_{\perp},
\label{3}
\end{equation}
the radiated power can be written as

\begin{equation}
P = \frac {2}{3} \frac {r_{0}c}{E_{0}}(\dot p_{||}^{2} +
\gamma^{2}p^{2}_{\perp}).
\label{4}
\end{equation}
where $E_{0} = mc^{2}$.
Equation (4) shows that the same acceleration force produces a $\gamma^{2}$
times higher radiation power, if it is applied in perpendicular direction,
compared to the parallel direction.

For a particle moving with a constant velocity in a circular machine
with bending radius $\varrho$ the power radiated due to curvature of the
orbit is

\begin{equation}
P = \frac {2}{3}r_{0}E_{0}c\frac {\beta^{4}\gamma^{4}}{\varrho^{2}}.
\label{5}
\end{equation}
So in the linear accelerator the energy loss caused by radiation
is smaller than in circle accelerator and it means that to obtain high
energy particles in linear accelerator is more easy than in
the circle accelerator.
In case of laser acceleration the situation radically changes .
The classical idea of laser acceleration is to consider the laser light
as the periodic electromagnetic field. The motion of electron in such a
wave was firstly described by Volkov (Berestetzkii et al. 1989).
However, it is possible to show that periodic electromagnetic wave does not
accelerate electrons in classical and quantum theory, because the
electric and magnetic components of the light field are mutually
perpendicular and it means the motion caused by the classical
periodic electromagnetic field is not linear but periodic (Landau et al.,
1962).

The situation changes if we consider laser beam as a system of
photons and the interaction of electron with laser light
is via the Compton process

\begin{equation}
\gamma + e \rightarrow \gamma + e.
\label{6}
\end{equation}

We can see that the right side of equation (6) involves no bremsstrahlung
photons and it means that there are
no energy loss caused by emission of photons. It means also
that laser acceleration is more effective than the acceleration in
the standard linear and circle accelerators.
It is evident that acceleration by laser can be adequately
described only by quantum field theory.
Such viewpoint gives us the motivation to
investigate theoretically the effectiveness of acceleration of charged
particles by laser beam.

\section{Historical view on laser acceleration}

The acceleration of charged particles by laser beam has been studied by many
authors (Tajima and Dawson, 1979; Katsouleas and Dawson, 1983;
Scully and Zubary, 1991; Baranova and Zel'dovich, 1994).

Many designs for such devices has been proposed.
Some of these are not sufficiently developed to be readily intelligible,
others seem to be fallacious and others are unlikely to be relevant to
ultra high energies. Some designs were developed only
to observe pressure of laser light on microparticles in liquids and gas
(Ashkin, 1970;  Ashkin, 1972).

The idea of laser acceleration follows historically
the idea that light exerts pressure. The former idea  was for the
first time postulated
by Johanness Kepler, the King astronomer in Prague, in 1619. He wrote that the
pressure of Sun light is what causes the tails of comets
to point away from the Sun.
The easy explanation of that effect was given by Newton in his corpuscular
theory of light where this effect is evidently of the mechanical origin.
Nevertheless, the numerical value of the light pressure was not known from
time of Kepler to the formulation of the Maxwell theory of electromagnetism
where Maxwell predicted in 1873 the magnitude of the light pressure.

The experimental terrestrial verification of the light pressure was given
by the Russian physicist Lebedev and Nichols and Hull from USA
(Nichols and Hull, 1903).The
measurement consisted in determination of force acting on the torsion
pendulum. At these experiments it was observed that the magnitude
of the pressure of light confirmed the Maxwell prediction.
It was confirmed that the pressure
is very small and practically has no meaning if the  weak terrestrial
sources of light are used. Only after invention of lasers the situation
changed
because of the very strong intensity of the laser light which can cause the
great pressure of the laser ray on the surface of the condensed matter.
So, the problem of the determination of the light pressure is now physically
meaningful because of the existence of high intensity lasers.

Here, we consider the acceleration of an
electron by laser beam.
We calculate the force due to Compton scattering of laser beam photons
on electron using the methods of
quantum field theory and quantum electrodynamics.

\section{Quantum field theory of a laser beam acceleration}

The dynamical equation of the relativistic particle with
rest mass $m$ and the kinematical mass $m(v)$,

\begin{equation}
m(v) = \frac {m}{\sqrt{1-\frac {v^{2}}{c^{2}}}},
\label{7}
\end{equation}
is as follows (M$\o$ller, 1972):

$${\bf F} = \frac {d(m(v){\bf v})}{dt} = m(v)\frac {d{\bf v}}{dt} +
\frac {dm(v)}{dt}{\bf v}
= m(v)\frac {d{\bf v}}{dt} + \frac {1}{c^{2}}\frac {dW}{dt}{\bf v} = $$

\begin{equation}
m(v)\frac {d{\bf v}}{dt} + \left(\frac {{\bf F}\cdot{\bf v}}{c^{2}}\right)
{\bf v},
\label{8}
\end{equation}
or,

\begin{equation}
m(v)\frac {d{\bf v}}{dt} = {\bf F} - \frac {{\bf v}}{c^{2}}({\bf F}\cdot
{\bf v}).
\label{9}
\end{equation}

If we consider a particle that is acted upon by a force ${\bf F}$
and which has an initial velocity in the direction of the force, then,
according to \eqnref{9} the particle will continue to move in the
direction of force. Therefore the path of the particle will be a straight
line, and we can choose this line as the $x$-axis. From \eqnref{9}
then follows for ${\bf v}\parallel {\bf F}$, and $F = |{\bf F}|$:

\begin{equation}
\frac {d}{dt}\left\{
\frac {v}{(1-\frac {v^{2}}{c^{2}})^{1/2}}\right\} = \frac {F}{m},
\label{10}
\end{equation}
where force $F$, in case it is generated by the laser beam,
contains also the velocity $v$ of particle as an integral part of the
Doppler frequency.

Now, let us consider the laser acceleration of an electron by the
monochromatic laser beam.
The force of photons acting on an electron in a laser beam depends evidently
on the density of photons in this beam. Using the definition of the
cross section of the electron-photon interaction and with the
energy loss $\omega-\omega'$, it may be easy to
define the force acting on electron by the laser beam, in the following
way:

\begin{equation}
F = n\int_{\omega_{1}}^{\omega_{2}}(\omega - \omega')
\frac {d\sigma(\omega -\omega')}{d(\omega -\omega')}
d(\omega -\omega'),
\label{force}
\end{equation}
where in the rest system of electron we have the following
integral $\omega'$-limits (Sokolov et al., 1983):
\begin{equation}
\omega_{2} = \frac {\omega}{1 + \frac {2\omega}{m}} \leq \omega'
\leq \omega = \omega_{1}.
\label{limits}
\end{equation}
where  $\omega$ can be identified with the frequency of the impinging
photon on the rest electron.

Let us remark that \eqnref{force} has the dimensionality of force
if we correctly suppose that the dimensionality of $d\sigma$ is
${\rm m}^{2}$ and density of photons in laser beam is ${\rm m}^{-3}$.
At the same time
the combination $\omega-\omega'$ in the cross section and differential
can be replaced by $\omega'$.

Since the
dimensionality of the expression on the right side of the last equation
is force, we at this moment connect $m$ with $c^{2}$ in order to get
$mc^{2} = E$. Quantity $\omega$ will be later denoted as the frequency
$\omega_{0}$ of the impinging photon. Using the expression for the
differential Compton cross section (Berestetzkii et al., 1989),
we get for the force:

\begin{equation}
F =  n\pi r_{e}^{2} \frac {E}{\omega^{2}}
\int_{\omega_{1}}^{\omega_{2}}d\omega'
(\omega'-\omega)
\left[\frac {\omega}{\omega'} + \frac {\omega'}{\omega} +
\left(\frac {m}{\omega'} - \frac {m}{\omega}\right)^{2} -
2m\left(\frac {1}{\omega'} - \frac {1}{\omega}\right)\right],
\label{force2}
\end{equation}
where $r_{e} = e^{2}/{m c^{2}}$ is the classical radius of an electron.

After $\omega'$-integration we obtain

\begin{eqnarray}
F = n\pi r_{e}^{2}\frac {E}{\omega^{2}}
\left\{(3m^{2} - \omega^{2} + 2m\omega)\ln\frac {\omega_{2}}{\omega_{1}} +
m^{2}\omega\left(\frac {1}{\omega_{2}}- \frac {1}{\omega_{1}}\right)
+ \right. \nonumber \\
\left. \left(\omega - 4m - \frac {3m^{2}}{_{\omega}}\right)(\omega_{2} -
\omega_{1}) + \frac {1}{3\omega}\left(\omega_{2}^{3} - \omega_{1}^{3}\right) +
\left(\frac {m^{2}}{2\omega^{2}} + \frac {m}{\omega} - \frac {1}{2}\right)
(\omega_{2}^{2}-\omega_{1}^{2})\right\},
\label{force3}
\end{eqnarray}
where the corresponding the $\omega_{i}$-combination are as follows:
\begin{equation}
\frac {\omega_{2}}{\omega_{1}} = \frac {m}{m + 2\omega};\quad
\omega_{2}-\omega_{1} = -\frac {2\omega^{2}}{m+2\omega};\quad
\frac {1}{\omega_{2}}-\frac {1}{\omega_{1}} = \frac {2}{m}
\label{omega1}
\end{equation}
and
\begin{equation}
\omega_{2}^{2}-\omega_{1}^{2} = -\frac
{4\omega^{3}(m+\omega)}{(m+2\omega)^{2}};\quad
\omega_{2}^{3}-\omega_{1}^{3} = -\frac {2\omega^{4}}{(m+2\omega)^{3}}
(3m^{2}+6m\omega+4\omega^{2}).
\label{omega2}
\end{equation}

Then, after insertion of the
$\omega_{i}$-combinations and putting  $\omega \rightarrow  \omega_{0}$,
we obtain for the accelerating force
instead of equation \eqnref{force3} the following equation:

\begin{eqnarray}
F =  n\pi r_{e}^{2}\frac {E}{\omega_{0}^{2}} \times \nonumber \\
\left\{(3m^{2} - \omega_{0}^{2} + 2m\omega_{0})
\ln\frac {m}{m + 2\omega_{0}} + 2m\omega_{0} +
\frac {2\omega_{0}(4m\omega_{0} + 3m^{2} - \omega_{0}^{2})}{m+2\omega_{0}}
\right. \nonumber \\
\left. - \frac {2\omega_{0}^{3}(3m^{2}+ 6m\omega_{0} +4\omega_{0}^{2})}
{3(m+2\omega_{0})^{3}} +
2\omega_{0}(\omega_{0}^{2}-m^{2}-2m\omega_{0})\frac {(m+\omega_{0})}
{(m+2\omega_{0})^{2}}\right\}.
\label{force4}
\end{eqnarray}

Now, if we want to express the force $F$ in the MKS system where its
dimensionality is kg.m$^{2}$ s$^{-2}$, we are forced to introduce the physical
constants; velocity of light c, Planck constant $\hbar$
in the last formula. It is easy to see that the last
formula expressed in the MKS system is as follows:

$$F = n\pi r_{e}^{2}\frac {E}{(\hbar \omega_{0})^{2}} \quad \times $$

$$\left\{(3m^{2}c^{4} - \hbar^{2}\omega_{0}^{2} + 2mc^{2}\hbar\omega_{0})
\ln\frac {mc^{2}}{mc^{2} + 2\hbar\omega_{0}} + 2mc^{2}\hbar\omega_{0}\quad +
\right.$$

$$\frac {2\hbar\omega_{0}(4mc^{2}\hbar\omega_{0} + 3m^{2}c^{4}
- \hbar^{2}\omega_{0}^{2})}{mc^{2}+2\hbar\omega_{0}} $$

$$- \quad \frac {2\hbar^{3}\omega_{0}^{3}(3m^{2}c^{4}+
6mc^{2}\hbar\omega_{0} +4\hbar^{2}\omega_{0}^{2})}
{3(mc^{2}+2\hbar\omega_{0})^{3}} + $$

\begin{equation}
\left. 2\hbar\omega_{0}(\hbar^{2}\omega_{0}^{2}-m^{2}c^{4} -
2mc^{2}\hbar\omega_{0})
\frac {(mc^{2}+\hbar\omega_{0})}
{(mc^{2}+2\hbar\omega_{0})^{2}}\right\}.
\label{force5}
\end{equation}

The last formula is valid approximately only for
nonrelativistic velocities because for laser photons the Doppler effect
plays substantional role for moving electron in the laser field.
The formula of the relativistic Doppler effect is as follows:

\begin{equation}
\omega_{0}\quad\longrightarrow\quad \omega_{0}\frac {1-\frac {v}{c}}
{\left(1-\frac {v^{2}}{c^{2}}\right)^{1/2}},
\label{Doppler}
\end{equation}
which means that for ultrarelativistic electron the frequency of photons
accelerating the electron will be very small and that the acceleration
will be also very small with regard to the electron moving with the
relativistic velocities in the laser field. In order to obtain the exact
description of the electron motion in the laser field, it is
necessary to insert the Doppler frequency equation \eqnref{Doppler}
in the formula \eqnref{10}.

Of course, it is not easy to obtain the general solution of \eqnref{10}
because it is strongly nonlinear.
For $v \ll c$ we obtain the solution $(E = mc^{2}, \hbar\omega_{0} =
\varepsilon)$:

$$v = \frac {1}{m}\pi n t r_{e}^{2} \quad \times $$

$$\left\{(3E^{2} - \varepsilon^{2} + 2E\varepsilon)
\ln\frac {E}{E + 2\varepsilon} + 2E\varepsilon \quad +
\right. $$

$$\frac {2\varepsilon(4E\varepsilon + 3E^{2} - \varepsilon^{2})}
{E + 2\varepsilon}\quad - \quad \frac {2\varepsilon^{3}(3E^{2} +
6E\varepsilon + 4\varepsilon^{2})}
{3(E + 2\varepsilon)^{3}}\quad + $$

\begin{equation}
\left. 2\varepsilon(\varepsilon^{2} - E^{2} - 2E\varepsilon)
\frac {(E + \varepsilon)}
{(E + 2\varepsilon)^{2}}\right\}.
\label{velocity2}
\end{equation}

We simplify the last result using the approximation $\varepsilon \ll E$. In
this approximation we use with $x = 2\varepsilon/E, \ln (1+x)^{-1}
\approx (-x + x^{2}/2 -x^{3}/3 + x^{4}/4), (1+x)^{-1} \approx
1-x+x^{2}-x^{3}, (1+x)^{-2} \approx 1-2x+3x^{2}-4x^{3}, (1+x)^{3}\approx
1-3x +6x^{2}$, and we obtain:

\begin{equation}
v \approx \frac {1}{m}
 n t \left(\frac {8\pi}{3}r_{e}^{2}\right)
\frac {\varepsilon^{2}}{E} = \frac {1}{m} n t \sigma_{\gamma e}
\frac {\varepsilon^{2}}{E}.
\label{velocity3}
\end{equation}
where $\sigma_{\gamma e}$ is the Thompson cross section.
It is evident that the general formula of the
velocity can be obtained in the form

\begin{equation}
v = \frac {1}{m} n t \sigma_{\gamma e} \varepsilon
f\left(\frac {\varepsilon}{E}\right)
\label{velocity4}
\end{equation}
where Taylor coefficients of $f$ can be obtained by expansion of
\eqnref{velocity2}.

Now, let us calculate the laser acceleration of an electron by the
monochromatic $2MW$ laser i. e.
with photon energy $\varepsilon = \hbar \omega_{0} =
2\times 10^{-20}$ J and density of photons $n = N/V =
4\times 10^{24} {\rm m^{-3}}$. Let us determine the velocity of accelerated
electron at $t= 1 {\rm s}$.
We use the following physical constants:
$m = 9.1 \times 10^{-31}{\rm kg}$, $c = 3 \times 10^{8}
{\rm m}.{\rm s}^{-1}$, $\sigma_{\gamma e} =
6.65 \times 10^{-29}{\rm m}^{2}$ (Muirhead, 1968).

After some numerical calculation we obtain the following numerical value of
velocity of electron
at time $t = 1 {\rm s}$:

$$v(t= 1 {\rm s}) \approx
\frac {1}{9.1 \times 10^{-31}} \times
4\times 10^{24}\times (2\times 10^{-20})^{2}
\times 6.65\times 10^{-29}\times $$

\begin{equation}
\frac {1}{9.1 \times 10^{-31}}\times \frac {1}{(3\times 10^{8})} \times 1
\approx 1.4 \; {\rm m}.{\rm s}^{-1}.
\label{velocity5}
\end{equation}

So, we see that for very short time intervals the motion of an
electron is nonrelativistic. For large time intervals
it will be necessary to use the relativistic equation involving
the relativistic mass and the Doppler effect.
The derived velocity, as we can see, is very small. However it can be
increased by increasing the the density of photons in a laser beam.
The density of photons
can be,  for instance, increased  by an appropriate focusation
of a laser beam. So, by this way, we can achieve the sufficient velocity of
electrons for the practical application.

\section{Discussion}

We have seen, in this article,  that the force accelerating an electron
by a laser beam can be determined by means of
the quantum field theory. The derived formula \eqnref{10} with
\eqnref{Doppler}
describes the force due to the Compton scattering of photons with
electron moving in the laser monochromatic photon sea. We have used only the
simple Compton process \eqnref{6} and not the more complicated
multiple Compton process defined by equation

\begin{equation}
n\gamma + e \quad \rightarrow \quad \gamma + e,
\label{ncompton}
\end{equation}
which follows for instance as a quantization of the Volkov
equation (Berestetzkii
et al., 1989).

The present article is the modification of the Pardy
discussion on laser acceleration
(Pardy, 1998), where the thermal statistical model of laser acceleration
was proposed. The basic ansatz of that model was the energy loss formula

\begin{equation}
-\frac {dW}{dx} = \frac {1}{v}\int d\Gamma (\omega - \omega'),
\label{eloss}
\end{equation}
where $\Gamma$ is the differential reaction rate defined in different manner
in quark qluon plasma physics and in electrodynamical medium.

In that article, it was used the approach by Brown et al.
(Brown and Steinke, 1997; Brown, 1992; Sokolov et al., 1983, Braaten
and Thoma, 1991).
At the same time we used the ideas of Blumenthal et al.
(Blumenthal and Gould, 1970). Brown et al. (Brown and Steinke, 1997),
applied the total electron scattering rate for determination of
behaviour of
electron in the Planckian photon sea inside of the pipe of the storage rings.
While in the preceding article the thermal distribution function $f(k)$
of photons was considered, here,
we used the nonthermal density of photons n.

The experimental perspective of the laser beam acceleration of elementary
particles concerns not only the charged particles, however, also
the neutral particles such as neutron, neutral $\pi$-meson,
and so on. Also the system of particles with the opposite charges was
considered to be
simultaneously accelerated by the laser beam (Pardy, 1997).
New experiments can be realized and new measurements performed by means of
the laser accelerator, giving new results and discoveries.
So, it is obvious that the acceleration of particles by the laser beam
can form, in the near future, the integral part of the particle physics.
In such laboratories as ESRF, CERN, DESY, SLAC and so on,
there is no problem to
install lasers with the sufficient power of the photon beam, giving
opportunity to construct the laser accelerator.

To say the final words, we hope, the
ideas of the present article open the way to laser accelerators and
will be considered as the integral part of the today particle physics.

\vspace{10mm}

\noindent
{\bf REFERENCES}

\vspace{7mm}

\noindent
Ashkin, A., (1970) Acceleration and trapping of particles by radiation
pressure, Phys. Rev. Lett. 24, 156-159.\\[2mm]
Ashkin, A., (1972) The pressure of laser light, Scientific
American {226} (2), 63-71.\\[2mm]
Baranova, N. B. and  Zel'dovich, B. Ya., (1994)
Acceleration of charged particles by laser beams,  JETP 78, (3), 249-258.
\\[2mm]
Berestetzkii, B. B.,  Lifshitz, E. M. and Pitaevskii L. P., (1989) Quantum
electrodynamics, (Moscow, Nauka). \\[2mm]
Blumenthal, G. R., and  Gould, R. J., (1970) Bremsstrahlung, synchrotron
radiation, and Compton Scattering of High-Energy Electrons
Traversing Dilute Gases, Rev. Mod. Phys. 42,  237-269.\\[2mm]
Braaten, E. and Thoma, M. H., (1991)  Energy loss of a heavy
fermion in a hot QED plasma, Phys. Rev. D 44, No. 4, 1298-1310.\\[2mm]
Brown, L. S., (1992) Quantum Field Theory, (Cambridge Univ. Press).\\[2mm]
Brown, L. S. and  Steinke, R. S., (1997),  Compton scattering on black
body photons,  Am. J. Phys. 65 (4), 304-309.\\[2mm]
Katsouleas, T. and Dawson, J. M.,  (1983) Unlimited electron
acceleration in laser-driven plasma waves,  Phys. Rev. Lett. 51, 392-395.
\\[2mm]
Landau, L. D.,  and Lifshitz, E. M., (1962) The Classical
Theory of Fields, 2nd ed.~(Pergamon Press, Oxford). \\[2mm]
Maier, M., (1991) Synchrotron radiation, in:
CAS-PROCEEDINGS, ed. S. Turner, CERN 91-04, 8 May, 97-115.\\[2mm]
M$\o$ller, C., (1972) The Theory of Relativity, (2nd edition),
(Clarendon Press, Oxford).\\[2mm]
Muirhead, H., (1968) The Physics of Elementary Particles, 2nd ed.
(Pergamon Press, Oxford, London). \\[2mm]
Nichols, E. F. and Hull, G. F., (1903) Phys. Rev. 17, 26-50; ibid. 91-104.
\\[2mm]
Pardy, M., (1997) \v Cerenkov effect and the Lorentz Contraction,
Phys. Rev. A 55, No. 3 , 1647-1652.\\[2mm]
Pardy, M., (1998) The quantum field theory of laser acceleration,
Phys. Lett. A 243, 223-228. \\[2mm]
Scully,  M. O. and  Zubary, M. S., (1991) Simple laser accelerator:
Optics and particle dynamics,  Phys. Rev. A 44 , 2656-2663.\\[2mm]
Sokolov, A. A., Ternov, I. M. ,  \v Zukovskii, V. \v C.  and
Borisov, A. B., (1983) Quantum Electrodynamics, (Moscow Univ. Press)
(in Russian).\\ [2mm]
Tajima, T. and Dawson, J. M., (1979) Laser Electron Accelerator,
Phys. Rev. Lett. 43, 267-270.

\end{document}